\documentclass[aps, prb, twocolumn, amssymb, amsmath, showpacs, superscriptaddress]{revtex4-1}
\usepackage{bm}
\usepackage{times}
\usepackage{graphicx}
\usepackage{color}
\usepackage{dcolumn}
\usepackage[colorlinks=true, letterpaper=true, pdfstartview=FitV, linkcolor=blue, citecolor=blue, urlcolor=blue]{hyperref}
\usepackage{appendix}
\usepackage{placeins}
\usepackage[normalem]{ulem}

\newcommand{\Tr}{\operatorname{Tr}}
\newcommand{\Ree}{\operatorname{Re}}

\begin{document}

\title{Andreev reflection mediated by topological corner states in a two-dimensional honeycomb lattice}

\author{Kai-Tong Wang}
\affiliation{School of Physics and Engineering, Henan University of Science and Technology, Luoyang 471023, China}

\author{Yunjin Yu}
\affiliation{College of Physics and Optoelectronic Engineering, Shenzhen University, Shenzhen 518060, China}

\author{Fuming Xu}
\email[]{xufuming@szu.edu.cn}
\affiliation{College of Physics and Optoelectronic Engineering, Shenzhen University, Shenzhen 518060, China}
\affiliation{Quantum Science Center of Guangdong-Hongkong-Macao Greater Bay Area, Shenzhen 518045, China}

\author{Jian Wang}
\email[]{jianwang@hku.hk}
\affiliation{College of Physics and Optoelectronic Engineering, Shenzhen University, Shenzhen 518060, China}
\affiliation{Quantum Science Center of Guangdong-Hongkong-Macao Greater Bay Area, Shenzhen 518045, China}
\affiliation{Department of Physics, The University of Hong Kong, Pokfulam Road, Hong Kong, China}

\begin{abstract}
Topological corner states in two-dimensional second-order topological insulators (SOTIs) are localized in real space. We numerically demonstrate that such localized topological corner states can mediate Andreev reflection when coupled to a superconducting lead. We consider a transport setup based on a two-dimensional honeycomb lattice, consisting of a normal lead, a central SOTI region, and a superconducting lead. The central SOTI region is described by the modified Kane--Mele model with an in-plane Zeeman field and hosts topological corner states in a diamond-shaped flake. Although the central region is insulating, the local density of states shows that incident electrons can turn the localized corner state into an extended scattering state, which forms a resonant tunneling channel to the superconducting lead. This process leads to a perfect Andreev reflection peak near zero energy. Away from this resonance, antiresonance dips appear in the Andreev reflection spectrum, and their positions can be tuned by the Zeeman field strength. We show that the suppression of Andreev reflection is caused by quantum interference and the imbalance between electron and hole dwell times in the central region. These results demonstrate that topological corner states can provide a resonant tunneling path to the superconducting interface and mediate Andreev reflection in second-order topological systems.
\end{abstract}
\maketitle

\section{INTRODUCTION}

Andreev reflection is a fundamental transport process occurring at the interface between a superconductor and another material~\cite{andreev64JETP}, in which an incident electron is reflected as a hole while a Cooper pair is injected into the superconductor. This electron-to-hole conversion depends sensitively on the available low-energy states and scattering channels near the superconducting interface, and hence Andreev reflection provides an effective probe of electronic states near such interfaces. Andreev reflection has been discussed in superconductors coupled to metals and quantum dots~\cite{marmorkos93PRB,sun99PRB}, ferromagnets~\cite{costa19PRB}, graphene~\cite{beenakker06PRL,beenakker08RMP,sun09JPCM}, topological insulators~\cite{sun11PRB,Narayan12PRB,Knez12PRL,hou16PRB,wei20PRB}, and various topological or nodal systems~\cite{diez12PRB,Bovenzi17PRB,cheng20PRB,azizi20PRB,yu18PRL,wei25PRL}. Recently, Andreev reflection has also been proposed in altermagnets, where spin-split bands can generate spin-polarized specular Andreev reflection~\cite{sun23Altermagnet,nagae25Altermagnet}.

Higher-order topological insulators (TIs) provide new types of topological boundary states. Conventional TIs are usually manifested by symmetry-protected conducting edge states or surface states~\cite{hasan10RMP,qi11RMP}. By contrast, higher-order TIs are characterized by insulating bulk states and lower-dimensional hinge modes or corner states appearing at selected boundaries~\cite{benalcazar17Science,benalcazar17PRB,schindler18SA,biye21NRP,hossain24NP}. In particular, second-order TIs (SOTIs) in two-dimensional (2D) systems are characterized by in-gap topological corner states localized in real space. Such topological corner states have been predicted in graphene-based honeycomb lattices~\cite{ren20PRL}, coupled 2D TIs~\cite{liu25PRB}, kagome lattices~\cite{ezawa18PRL}, graphdiyne~\cite{sheng19PRL}, C$_2$N and its derivatives~\cite{li22PRB}, twisted bilayer graphene~\cite{park19PRL,park21carbon}, quasicrystals~\cite{rui20PRL}, topological Anderson systems~\cite{zhang21PRL,Yang21PRB}, transition-metal dichalcogenides~\cite{wang19PRL,zeng21PRB}, and moir\'e superlattices~\cite{Liu21PRL}.

Since topological corner states are localized, they do not form ordinary transport channels by themselves. Previous works have shown that, in 2D honeycomb lattices, incoming electrons can transform localized corner states into extended resonant states and induce resonant tunneling processes mediated by topological corner states~\cite{ktwang21SCPMA,ktwang22FOP}. In other SOTIs, transport through topological corner states has also been utilized to realize quantized charge pumping~\cite{wu22PRB}, quantized spin pumping~\cite{long23SCPMA}, spin filtering~\cite{ktwang24PRB}, and topological nanoswitching~\cite{poata24NJP}. Although Andreev reflection has been extensively studied in TI--superconductor junctions, whether topological corner states can mediate Andreev reflection through an insulating SOTI remains unclear.

In this work, we numerically investigate this problem based on the modified Kane--Mele model for a 2D honeycomb lattice with an in-plane Zeeman field~\cite{ren20PRL}. The two-terminal transport setup consists of a left normal lead, a central SOTI region hosting a topological corner state, and a top superconducting lead [Fig.~\ref{fig1}(a)]. Since the central SOTI region is insulating, ordinary bulk transport between the two leads is forbidden. Nevertheless, the partial local density of states reveals that incident electrons can turn the localized corner state into an extended scattering state, which forms an effective transport path to the superconducting lead. As a result, a perfect Andreev reflection peak appears near zero energy. Away from this resonance, Andreev reflection is suppressed due to quantum interference and the imbalance between the electron and hole dwell times in the central region. We further demonstrate that the Zeeman field can control the positions of these antiresonance dips, providing an efficient way to tune Andreev reflection mediated by topological corner states.

The rest of this paper is organized as follows. In Sec.~II, we introduce the model Hamiltonians for the SOTI--superconductor hybrid system and the nonequilibrium Green's function formalism for transport calculation. In Sec.~III, we present numerical results and discuss the underlying mechanism of Andreev reflection mediated by topological corner states. Finally, a brief summary is given in Sec.~IV.

\bigskip
\section{MODEL AND FORMALISM}

\subsection{Model Hamiltonian}

\begin{figure}[t!]
\centering
\includegraphics[width=\columnwidth]{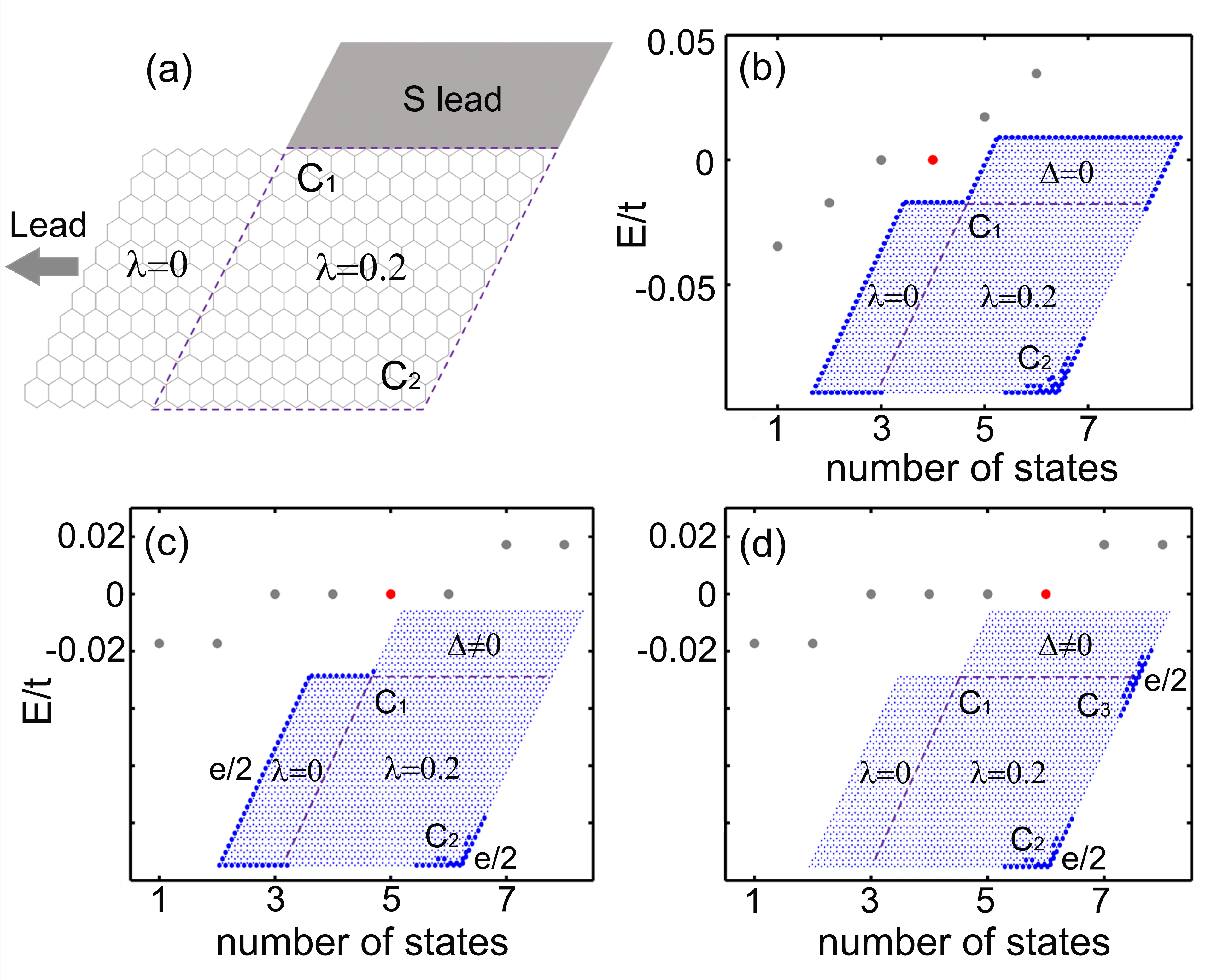}
\caption{(a) Schematic of the two-terminal SOTI-superconductor hybrid system, where the central diamond-shaped lattice is connected to a left normal lead and a top superconducting lead (S lead). (b) Energy levels of the isolated central scattering region with $\lambda=0.2$ and $\Delta=0$. The inset shows the eigenfunction distribution of the zero-energy state labeled by the red dot. (c)-(d) Energy levels and eigenfunction distributions of the fifth and sixth states near $E =0$, respectively, marked by red dots, for $\lambda=0.2$ and $\Delta \neq 0$. } \label{fig1}
\end{figure}

The model system is shown in Fig.~\ref{fig1}(a), which consists of a left normal lead, a central diamond-shaped second-order topological insulator (SOTI), and a top superconducting lead. The total Hamiltonian is written as
\begin{equation}\label{eq:total}
H = H_{L}+H_{C}+H_{LC}+H_{S}+H_{SC}.
\end{equation}
Here $H_L$ and $H_S$ describe the left normal lead and the top superconducting lead, respectively. $H_C$ is the Hamiltonian of the central SOTI region. $H_{LC}$ and $H_{SC}$ describe the couplings between the central region and the two leads.

We first introduce the central region, which hosts the topological corner state. We adopt the modified Kane--Mele model for a two-dimensional honeycomb lattice. In this model, an in-plane magnetic field breaks the time-reversal symmetry of the conventional Kane--Mele Hamiltonian and drives the system into a second-order topological phase with corner states at the obtuse corners of a diamond-shaped flake~\cite{ren20PRL,ktwang21SCPMA}. The Hamiltonian of this modified Kane-Mele model is expressed as
\begin{equation}\label{eq:hc}
H_C = H_0+\lambda \sum_i c_i^{\dag}\mathbf{B}\cdot\mathbf{s}c_i,
\end{equation}
where $H_0$ is the conventional Kane--Mele Hamiltonian~\cite{Kane05PRL},
\begin{equation}\label{eq:h0}
H_0=-t\sum_{\langle ij\rangle} c_i^{\dag}c_j
+it_{SO}\sum_{\langle\langle ij\rangle\rangle}
\nu_{ij}c_i^{\dag}s_zc_j.
\end{equation}
Here $c_i^{\dag}=[c_{i,\uparrow}^{\dag},c_{i,\downarrow}^{\dag}]$ is the creation operator at site $i$, and $t$ is the nearest-neighbor hopping energy. The second term in $H_0$ denotes the intrinsic spin-orbit coupling with strength $t_{SO}$, which is described by next-nearest-neighbor hopping. The factor $\nu_{ij}=\pm1$ depends on whether the electron takes a clockwise or anticlockwise turn from site $j$ to site $i$. The last term in $H_C$ is the Zeeman term induced by an in-plane magnetic field, where $\mathbf{s}$ is the vector of Pauli matrices for spin and $\lambda$ is the Zeeman field strength. Throughout this work, we set $t_{SO}=0.1t$. For an isolated diamond-shaped flake, the Zeeman field gaps the boundary edge states and leaves zero-energy corner states near the obtuse corners~\cite{ren20PRL}, such as the $C_2$ corner shown in the inset of Fig.~\ref{fig1}(b).

% (d) Energy levels and eigenfunction distribution of the $6^{th}$ state (red dot) near zero-energy for $\lambda=0.2$ and $\Delta \neq 0$.

The left lead is described by the conventional Kane--Mele Hamiltonian $H_L=H_0$, with $H_0$ given in Eq.~(\ref{eq:h0}). Since the Zeeman term is absent in the left lead, it supports quantum spin Hall edge states, as shown in Fig.~\ref{fig2}(a). The coupling between the left lead and the central region is
\begin{equation}\label{eq:normal-coupling}
H_{LC} = -t \sum_{\langle ij\rangle} d_i^{\dag} c_{j}
+ it_{SO}\sum_{\langle\langle ij\rangle\rangle}
\nu_{ij} d_i^{\dag} s_z c_j + H.c.,
\end{equation}
where $d_i^{\dag}$ is the creation operator in the left lead at site $i$.

The top lead has the same normal-state Kane--Mele Hamiltonian as the left lead and is proximitized by an $s$-wave superconductor. In the Nambu basis $(e\uparrow,e\downarrow,h\uparrow,h\downarrow)$, it is described by
\begin{equation}\label{eq:bdg}
\mathcal{H}_{S}^{\rm BdG}=
\begin{pmatrix}
H_0 & i\Delta s_y \\
-i\Delta s_y & -H_0^{*}
\end{pmatrix}.
\end{equation}
Here $H_0$ is the conventional Kane--Mele Hamiltonian in Eq.~(\ref{eq:h0}). $\Delta$ is the superconducting gap, and the off-diagonal blocks describe the spin-singlet superconducting pairing. Therefore, when $\Delta=0$, the top lead is reduced to the same Kane-Mele Hamiltonian $H_0$ as the left lead. In the numerical calculation, the top superconducting lead is included through its self-energy. The coupling between the central region and the top superconducting lead is~\cite{hou16PRB,wei20PRB}
\begin{equation}\label{eq:superconductor-coupling}
H_{SC} = -t \sum_{\langle ij\rangle} c_i^{\dag} b_j
+it_{SO}\sum_{\langle\langle ij\rangle\rangle}
\nu_{ij} c_i^{\dag}s_z b_j + H.c.,
\end{equation}
where $b_j$ is the annihilation operator in the top lead. This coupling has a similar form as that in Eq.~(\ref{eq:normal-coupling}), except that $b_j = \sum_{\mathbf{k}} e^{i \mathbf{k} \cdot \mathbf{r}_{j}} b_{\mathbf{k}}$ is the annihilation operator in the superconducting lead with the momentum $ \mathbf{k} = (k_{x},k_{y})$.

\smallskip
\subsection{Transport properties from the nonequilibrium Green's function formalism}

The transport properties are calculated in the Nambu representation $(e \uparrow, e \downarrow, h \uparrow, h \downarrow)$. The Andreev-reflection coefficient $T_A$ describes the probability that incident electrons from the left lead are reflected back as holes. Using the nonequilibrium Green's-function formalism, the spin-resolved Andreev-reflection coefficient is~\cite{sun09JPCM}
\begin{equation}\label{eq:ar}
T_{A} =  \Tr[\Gamma_{L\sigma} \mathbf{G}_{\sigma\overline{\sigma}}^{r} \Gamma_{L\overline{\sigma}} \mathbf{G}_{\overline{\sigma}\sigma}^{a}],
\end{equation}
where $\sigma=\uparrow,\downarrow$ and $\overline{\sigma}$ denotes the opposite spin. The block Green's function $\mathbf{G}_{\sigma\overline{\sigma}}^r$ connects the incident electron with spin $\sigma$ to the reflected hole with spin $\overline{\sigma}$. The retarded Green's function of the central scattering region is defined as~\cite{Datta1995}
\begin{equation}\label{eq:Gr}
\mathbf{G}^r=[E-{\cal H}_C-\Sigma_L^r-\Sigma_S^r]^{-1},
\end{equation}
and $G^a=G^{r\dag}$. Here ${\cal H}_C =[H_C, 0; 0, -H_C^{*}]$ is the Hamiltonian of the central SOTI region in the Nambu representation, where $H_C$ is given in Eq.~(\ref{eq:hc}). ${\cal H}_C$ is understood as the normal central Hamiltonian $H_C$ acting in the electron and hole sectors, without superconducting pairing in the central region. Then the block $\mathbf{G}_{\sigma\overline{\sigma}}^r$ can be extracted from $\mathbf{G}^r$ accordingly.

%Finally, the self-energy of the superconducting lead is $\Sigma_S^r=V_{SC}g_S^rV_{SC}^{\dag}$, where $V_{SC}$ is the coupling matrix corresponding to Eq.~(\ref{eq:superconductor-coupling}) and $g_S^r$ is the surface Green's function of the superconducting lead.
% , where $H_{SC}$ is the coupling matrix in Eq.~(\ref{eq:superconductor-coupling}) and $g_S^r$ is the surface Green's function of the superconducting lead, which is
The linewidth function of the left normal lead is defined as $\Gamma_{L}=i[\Sigma_{L}^{r}-\Sigma_{L}^{a}]$~\cite{Datta1995}, and the self-energy $\Sigma_L^r$ is calculated via the transfer-matrix method~\cite{Lee811PRB,Lee812PRB} using the Nambu form of the left-lead Hamiltonian ${\cal{H}}_L = [H_L, 0; 0, -H_L^{*}]$. The spin-resolved linewidth function $\Gamma_{L\sigma}$ can be extracted from $\Gamma_L$ in a similar way. The self-energy of the top superconducting lead  $\Sigma_S^r$ is defined as~\cite{xing04PRB,sun09JPCM,wei20PRB}
\begin{equation}\label{eq:gr}
\Sigma_S^{r}({r_i,r_j}) =  -i\pi t^2 \beta(E) J_{0}(k_{F}|r_i-r_j|)\left(\begin{smallmatrix}1&\Delta/E\\\Delta^{*}/E&1\end{smallmatrix}\right),
\end{equation}
where $r_i(r_j)$ represents the sites of the superconducting lead adjacent to the central region and $t$ is the nearest-neighbor hopping. For $|E|>\Delta$, $\beta(E) = |E|/\sqrt{E^{2}-\Delta^{2}}$, while for $|E|<\Delta$, $\beta(E) = E/(i\sqrt{\Delta^{2}-E^{2}})$. $J_{0}(k_{F}|r_i-r_j|)$ is the zeroth-order Bessel function and $k_{F}$ is the Fermi wave vector. % and $\rho$ denotes the density of states

To reveal the microscopic transport feature, we calculate the partial local density of states (PLDOS). This quantity describes how electronic states injected from the normal lead distribute inside the central scattering region and therefore gives a real-space picture of the Andreev-reflection process. Similar to the spin-resolved Andreev coefficient $T_A$ in Eq.~(\ref{eq:ar}), the electron PLDOS with spin $\sigma$ is expressed as
\begin{equation}\label{eq:ldos}
\rho_L^{e}(j,E)
=\frac{1}{2\pi}\Ree
\left[ \mathbf{G}_{\sigma\sigma}^r \Gamma_{L\sigma}
\mathbf{G}_{\sigma\sigma}^a \right]_{jj},
\end{equation}
while the hole PLDOS with opposite spin $\overline{\sigma}$ is
\begin{equation}
\rho_L^{h}(j,E)
=\frac{1}{2\pi} \Ree
\left[ \mathbf{G}_{\overline{\sigma}\sigma}^r
\Gamma_{L\sigma}
\mathbf{G}_{\sigma\overline{\sigma}}^a \right]_{jj}.
\end{equation}
Then the partial density of states is obtained by integrating the PLDOS over the central scattering region,
\begin{equation}\label{eq:partial-dos}
{\rm DOS}_{L}^{e/h}(E)=\sum_{j\in C}\rho_L^{e/h}(j,E).
\end{equation}

%% \begin{equation}\label{eq:ldos}
%\begin{aligned}
%\rho_L^{e}(j,E)
%&=\frac{1}{2\pi}\Ree
%\left[
%\mathbf{G}_{\sigma\sigma}^r
%\Gamma_{L\sigma}
%\mathbf{G}_{\sigma\sigma}^a
%\right]_{jj},\\
%\rho_L^{h}(j,E)
%&=\frac{1}{2\pi}\Ree
%\left[
%\mathbf{G}_{\overline{\sigma}\sigma}^r
%\Gamma_{L\sigma}
%\mathbf{G}_{\sigma\overline{\sigma}}^a
%\right]_{jj}.
%\end{aligned}
%\end{equation}

\bigskip
\section{NUMERICAL RESULTS AND DISCUSSION}

In numerical calculations, we consider the two-terminal transport setup shown in Fig.~\ref{fig1}(a). In this setup, a superconducting lead (S lead) is coupled to the top boundary of the central diamond-shaped lattice, while a normal lead with zero Zeeman field ($\lambda=0$) is attached to the left boundary. The SOTI region is indicated by the purple dashed rhombus. The widths of both the normal and superconducting leads are fixed at $W=30a$, where $a$ is the lattice constant. We take the hopping amplitude $t=2.75$ eV as the energy unit and set the superconducting gap to $\Delta = 0.0005t$.

\begin{figure}[t!]
\centering
\includegraphics[width=\columnwidth]{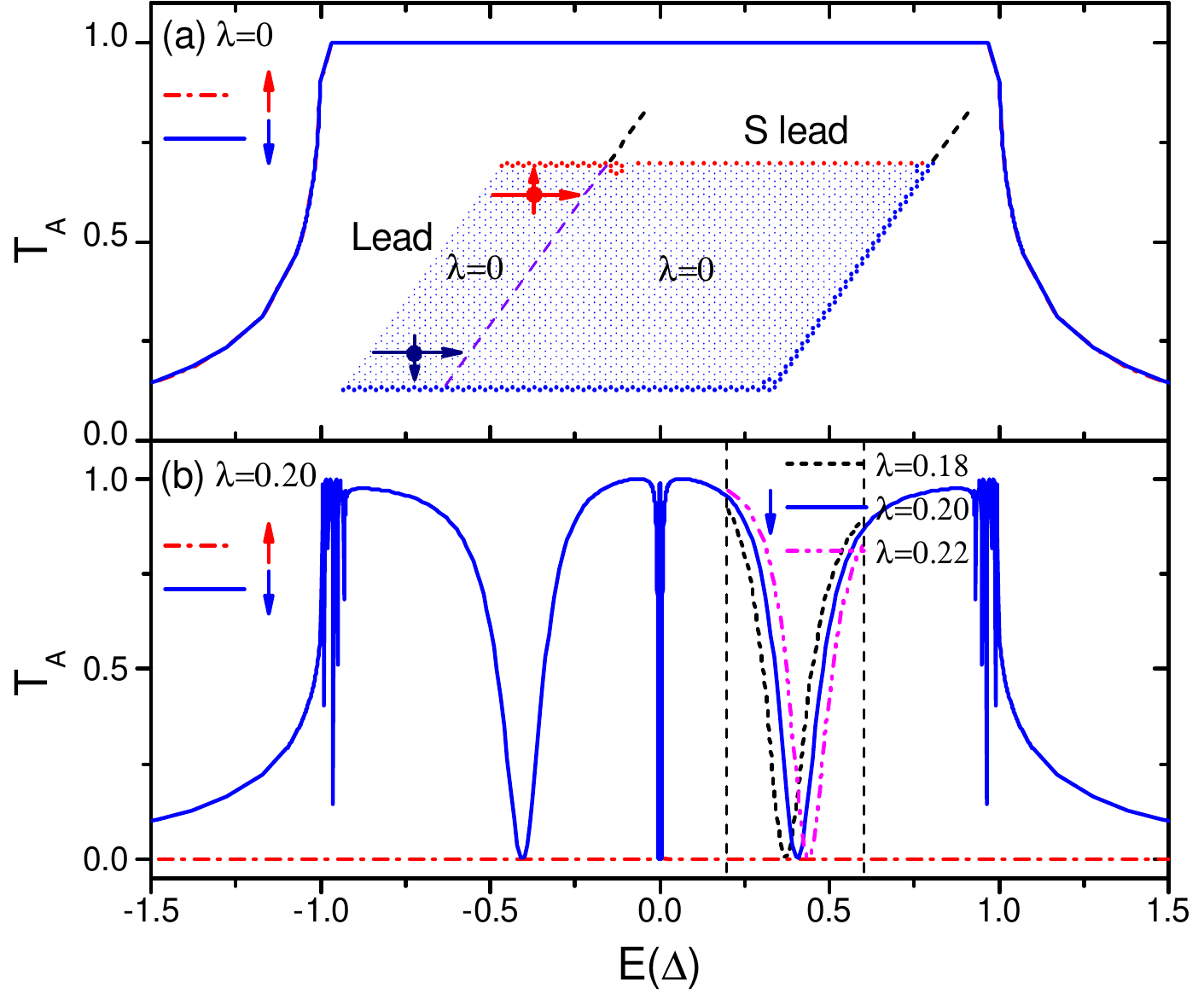}
\caption{(a) Andreev reflection coefficient $T_A$ as a function of the Fermi energy $E$ of incident electrons for $\lambda=0$. Inset: PLDOS distributions of the spin-up and spin-down edge states. (b) $T_A$ versus $E$ for $\lambda=0.2$. Inset: $T_A$ versus $E$ for different strengths of the Zeeman field in the region enclosed by the black dashed lines.}\label{fig2}
\end{figure}

We first examine the eigenfunction distributions of the isolated system, which provide the microscopic basis for Andreev reflection mediated by topological corner states. The isolated system consists of a diamond-shaped lattice together with finite segments of the left and top leads, as shown in the inset of Fig.~\ref{fig1}(b). In the absence of superconducting pairing ($\Delta = 0$), the corresponding energy levels are plotted in Fig.~\ref{fig1}(b). For the zero-energy state highlighted by the red dot, a single corner state resides near the $C_2$ corner, where both the zero-energy and localized features characterize topological corner states~\cite{ren20PRL}. Meanwhile, the edge states extend continuously along the zigzag boundaries of the $\lambda =0$ regions in Fig.~\ref{fig1}(b). When the superconducting pairing is turned on ($\Delta \neq 0$), four near-zero-energy states appear in the energy spectrum. As shown in Fig.~\ref{fig1}(c), the fifth state marked by the red dot retains a pronounced corner state near $C_2$. Importantly, this $C_2$ corner state carries a fractional charge of $e/2$, while the other $e/2$ charge is carried by the extended edge state in the left region. By contrast, Fig.~\ref{fig1}(d) shows that the sixth state has a different hybridized form: half of its weight(charge) remains localized near the $C_2$ corner, whereas the other half appears near the SOTI--superconductor interface around the $C_3$ corner. Thus the superconducting coupling converts the near-zero-energy states into an edge-corner state and a corner-interface hybrid state. These near zero-energy states provide the transport channels through which the insulating central region can support resonant tunneling.

\begin{figure}[tbp]
\centering
\includegraphics[width=\columnwidth]{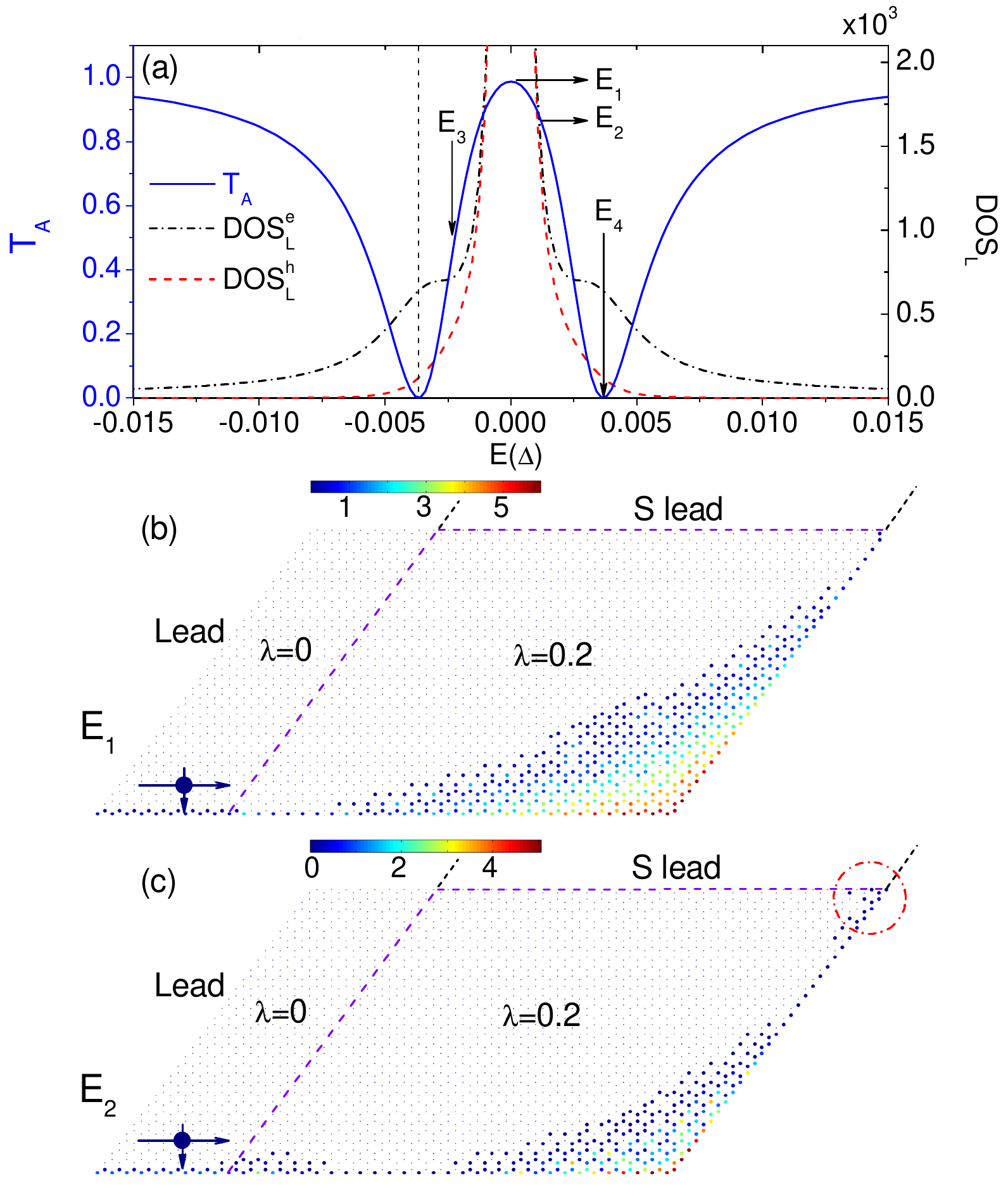}
\caption{(a) $T_A$, electron and hole partial densities of states ${\rm DOS}_L^e$ and ${\rm DOS}_L^h$ as functions of the Fermi energy $E$ near zero energy for $\lambda=0.2$. PLDOS distributions are shown for (b) $E_1=0$ and (c) $E_2=0.0012\Delta$, respectively.}\label{fig3}
\end{figure}

We next calculate the Andreev reflection coefficient $T_A$. As a reference, Fig.~\ref{fig2}(a) shows $T_A$ as a function of the Fermi energy $E$ for $\lambda=0$, where the central region is in the quantum spin Hall phase. Within the superconducting gap, the helical edge channels are efficiently coupled to the superconducting lead, and both spin-resolved channels undergo complete Andreev reflection. This gives rise to a quantized plateau with $T_A=1$. The inset shows the corresponding PLDOS distributions of the spin-up and spin-down edge states, which propagate along opposite boundaries and form helical transport channels connecting the normal and superconducting leads. When the in-plane Zeeman field is turned on, the transport channels are qualitatively reconstructed. Figure~\ref{fig2}(b) shows $T_A$ for $\lambda=0.2$. The spin-up channel is strongly suppressed, as indicated by the nearly zero red dashed curve, whereas the spin-down channel along the lower boundary remains active. For incident spin-down electrons, $T_A$ exhibits pronounced resonant and antiresonant behavior inside the superconducting gap. In particular, a near zero-energy resonance in $T_A$ is accompanied by pairs of finite-energy dips, where $T_A$ is strongly suppressed. The inset of Fig.~\ref{fig2}(b) further shows that the dip positions shift as $\lambda$ is varied. This behavior indicates that the suppression of $T_A$ originates from quantum interference between incident electrons and reflected holes, which depends on the Fermi energy and the Zeeman field. Therefore, changing $\lambda$ shifts the destructive-interference condition and moves the antiresonance dips. These features demonstrate that the Zeeman field not only selects the active spin channel, but also tunes the resonant tunneling process through the topological insulating region. The eigenstate analysis in Fig.~\ref{fig1} identifies the corner state and the corner-interface hybrid state near zero energy, while the PLDOS in Fig.~\ref{fig3} shows how these states are activated by incident electrons in the open transport setup.

Now we focus on the zero-energy resonance peak and the nearby antiresonance dips of Andreev reflection highlighted in Fig.~\ref{fig3}(a), and clarify their microscopic origin. Figure~\ref{fig3}(a) compares $T_A$ near zero energy with the electron and hole partial densities of states from the left lead, denoted by ${\rm DOS}_L^e$ and ${\rm DOS}_L^h$, respectively. The partial density of states (PDOS) is directly connected to an important dynamical transport property, the dwell time, which characterizes the time spent by a propagating electron or hole in the central scattering region. Following the Wigner--Smith delay time and the relation between PDOS and dwell time in the scattering matrix theory~\cite{wigner55PR,smith60PR,buttiker93PRL,iannaccone95PRB,gasparian96PRA,brouwer97EPL,xu11PRB,ossipov18PRL}, the dwell time can be defined as
\begin{equation}\label{eq:dwell-time}
\tau_d^{e/h}(E) = h{\rm DOS}_L^{e/h}(E).
\end{equation}
Here ${\rm DOS}_L^e$ and ${\rm DOS}_L^h$ are obtained by integrating the electron and hole PLDOS over the central scattering region. At the resonance ($E_1=0$), ${\rm DOS}_L^e$ is nearly equal to ${\rm DOS}_L^h$, so the electron and hole dwell times are balanced, leading to $T_A=1$. Near the antiresonance dips, the imbalance between the electron and hole dwell times reaches its maximum. Since the central SOTI region remains insulating, PLDOS is nearly zero away from the corner and interface regions. Therefore, the Andreev reflection process cannot be understood as ordinary bulk propagation from the normal lead to the superconducting lead. In the present system, it is instead mediated by resonant tunneling through scattering-extended topological corner states inside the insulating region.

This mechanism is directly revealed by the PLDOS distributions in Figs.~\ref{fig3}(b) and \ref{fig3}(c). At $E_1=0$, the incident electron is strongly coupled to the topological corner state near the $C_2$ corner. The scattering-extended corner state provides a resonant channel through the insulating central region and assists the conversion of incident electrons into reflected holes, giving rise to the resonant peak of $T_A = 1$. Similar resonant tunneling phenomena assisted by topological corner states have been extensively discussed in two-dimensional honeycomb lattices~\cite{ktwang21SCPMA, ktwang22FOP,ktwang24PRB}. At $E_2=0.0012\Delta$, the scattering state exhibits a hybridized character, with one component near the $C_2$ corner and another extending toward the SOTI--superconductor interface. This corner--interface hybrid state effectively connects the left normal lead to the top superconducting lead, allowing the incident electron to reach the pairing region and return as a hole with opposite momentum. Compared with Fig.~\ref{fig3}(b), the PLDOS in Fig.~\ref{fig3}(c) is less extended along the lower boundary, consistent with the slightly reduced Andreev reflection coefficient, $T_A(E_2)<T_A(E_1)$.

When the Fermi energy moves away from $E_1=0$, the Andreev reflection coefficient $T_A$ is gradually suppressed and eventually develops antiresonance dips. This suppression is closely related to the imbalance between the electron and hole partial densities of states. As shown in Fig.~\ref{fig3}(a), away from the zero-energy resonance, the electron PDOS ${\rm DOS}_L^{e}$ becomes larger than the hole PDOS ${\rm DOS}_L^{h}$. Consequently, the electron dwell time exceeds the hole dwell time, i.e., $\tau_d^e>\tau_d^h$. Such an imbalance implies that incident electrons are trapped in the central scattering region without being efficiently converted into holes, leading to the suppression of $T_A$. This imbalance reaches its maximum near the antiresonance dips, where Andreev reflection is completely suppressed. Therefore, these finite-energy dips originate from the competition between Andreev conversion mediated by the topological corner state and the imbalance in electron-hole dwell times.

\begin{figure}[tbp]
\centering
\includegraphics[width=\columnwidth]{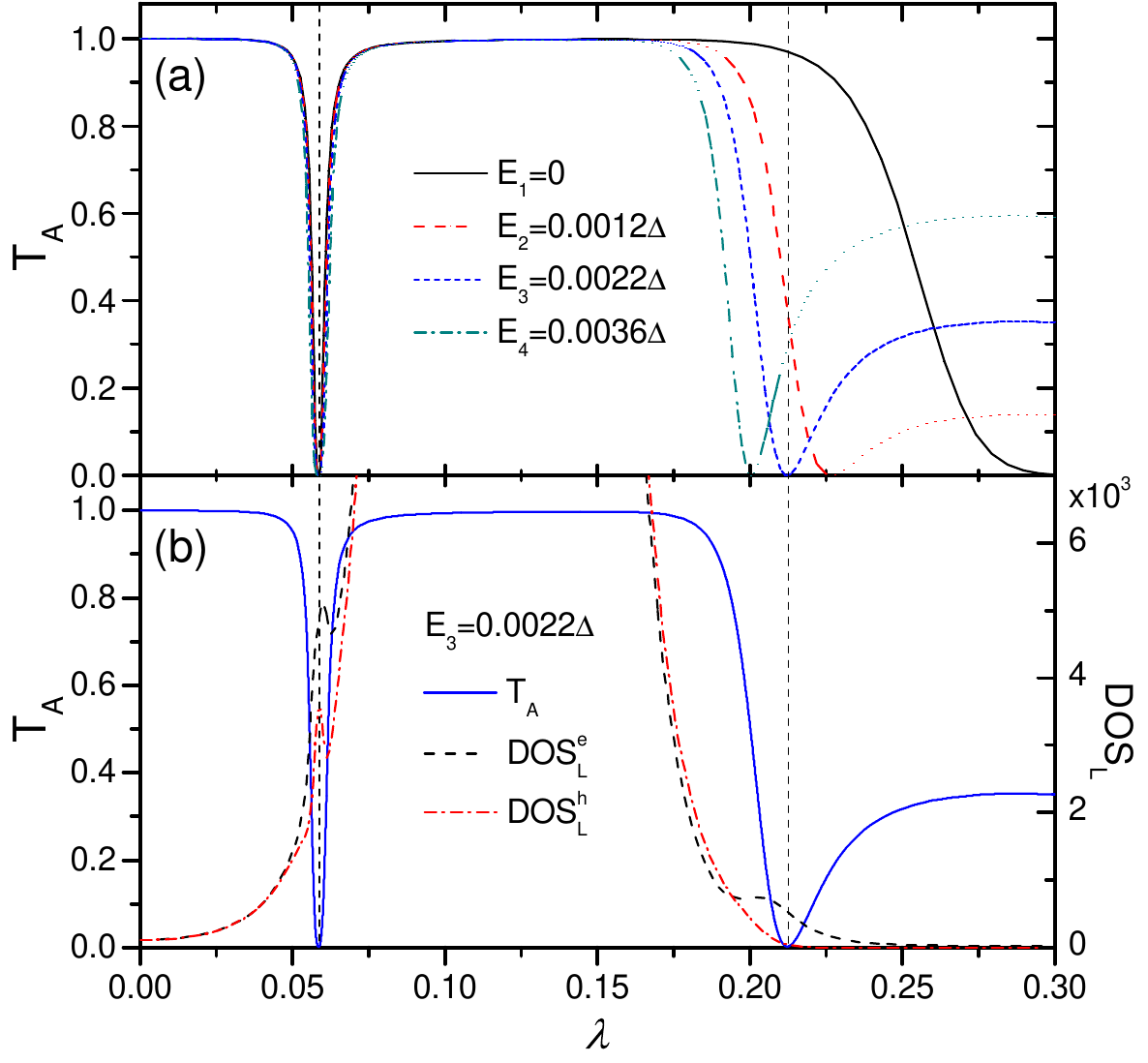}
\caption{(a) Andreev reflection coefficient $T_A$ as a function of the Zeeman field strength $\lambda$ for different Fermi energies. (b) $T_A$, ${\rm DOS}_L^e$, and ${\rm DOS}_L^h$ as functions of $\lambda$ for $E_3=0.0022\Delta$. The vertical dashed lines indicate antiresonance dips. }\label{fig4}
\end{figure}
% : $E_1=0$, $E_2=0.0012\Delta$, $E_3=0.0022\Delta$, and $E_4=0.0036\Delta$.
% The left vertical axis is for $T_A$, and the right vertical axis is for ${\rm DOS}_L^{e,h}$.

Finally, we examine how Andreev reflection is tuned by the Zeeman field strength. Figure~\ref{fig4}(a) shows $T_A$ as a function of $\lambda$ for different Fermi energies. For small $\lambda$, $T_A$ remains close to unity, except for a sharp antiresonance dip near $\lambda \simeq 0.052$ that appears for all four energies. The position of this first dip is nearly independent of energy. After this dip, $T_A$ forms a plateau close to unity over a broad range of $\lambda$. With further increasing $\lambda$, a second antiresonance dip appears around $\lambda \simeq 0.2$. In contrast to the first dip, the position of this second dip clearly depends on the Fermi energy. As $E$ is varied from $E_1$ to $E_4$, the dip shifts toward smaller $\lambda$. This energy dependence is a characteristic signature of quantum interference, since the destructive interference condition depends on both the quasiparticle energy and the Zeeman field strength.

To further clarify this behavior, Fig.~\ref{fig4}(b) shows $T_A$, ${\rm DOS}_L^e$, and ${\rm DOS}_L^h$ as functions of $\lambda$ for $E_3=0.0022\Delta$. The two vertical dashed lines mark the two antiresonance dips of $T_A$. Around these dips, ${\rm DOS}_L^e$ exceeds ${\rm DOS}_L^h$, indicating that the suppressed Andreev reflection is accompanied by dwell time imbalance when the dip positions are tuned by the Zeeman field strength. Thus, Fig.~\ref{fig4} confirms the mechanism inferred from Fig.~\ref{fig3}: the Zeeman field affects Andreev reflection by tuning both the transport path mediated by the topological corner state and the balance between electron and hole dwell times.
% Around these dips, the partial densities of states are strongly enhanced, indicating that carriers dwell for a long time in the central scattering region. %

\bigskip
\section{CONCLUSION}

In summary, we have numerically investigated Andreev reflection in a normal--SOTI--superconductor hybrid system realized on a two-dimensional honeycomb lattice. Although the central SOTI region is insulating, electrons incident from the left normal lead can turn the localized topological corner state into an extended scattering state and establish an effective transport path to the superconducting lead. The PLDOS distributions reveal that resonant tunneling channels are formed through the corner state and the hybridized corner-interface state near the SOTI--superconductor boundary, resulting in a perfect Andreev reflection peak near zero energy. The Zeeman field reconstructs the transport channels and tunes the quantum interference between incident electrons and reflected holes, giving rise to finite-energy antiresonance dips. Further analysis shows that the suppression of Andreev reflection originates from the imbalance between electron and hole dwell times in the central scattering region. These results demonstrate that topological corner states can provide a resonant tunneling path to the superconducting interface and mediate Andreev reflection in 2D second-order topological systems.

\bigskip
\section*{ACKNOWLEDGMENTS}

This work was supported by the National Natural Science Foundation of China (Grants No. 12304061 and No. 12574054). F.X. also acknowledges the Shenzhen Science and Technology Program (Grant No.~JCYJ2025).

\end{document}